\documentstyle [a4,epsfig]{article}
\def\mapbelow#1{\smash{\mathop{\longrightarrow}\limits_{#1}}}

\newcommand{\lab}{\label}

\newcommand{\bc}{\begin{center}}
\newcommand{\ec}{\end{center}}
\newcommand{\be}{\begin{equation}}
\newcommand{\ee}{\end{equation}}
\newcommand{\bea}{\begin{eqnarray}}
\newcommand{\eea}{\end{eqnarray}}
\newcommand{\bs}{\begin{subequations}}
\newcommand{\es}{\end{subequations}}
\newcommand{\beq}{\begin{eqalignno}}
\newcommand{\eeq}{\end{eqalignno}}

\newcommand{\half}{\frac{1}{2}}

\def\Journal#1#2#3#4{{#1} {\bf #2}, {#3} {(#4)}}

\def\PLA{{\em Phys. Lett.}  \bf A}

\def\al{\alpha}

\def\om{\omega}
\def\Om{\Omega}

\def\lab{\label}
\begin{document}

\thispagestyle{empty}

\vspace{2.0cm}
\bc

\huge{Formation and life-time of memory domains\\
in the dissipative quantum model of brain}  

\vspace{1.2cm}

\large{ Eleonora Alfinito$^{1,2}$, and Giuseppe Vitiello$^{1,3}$} \\
\small
{\it ${}^{1}$Dipartimento di Fisica, Universit\`a di Salerno, 84100 
Salerno, Italy\\
${}^{2}$INFM Sezione di Salerno}\\
${}^{3}$INFN Gruppo Collegato di Salerno\\
alfinito@sa.infn.it\\
vitiello@sa.infn.it

%{\it ${}^{\dag}$ OTHER}
\vspace{1.5cm}

\ec

\normalsize
{\bf Abstract}

We show that in the dissipative quantum model of brain the time-dependence 
of the frequencies of the electrical dipole wave quanta leads to the 
dynamical organization of the memories in space (i.e. to their 
localization in more or less diffused regions of the 
brain) and in time (i.e. to their longer or shorter life-time). The 
life-time and the localization in domains of the memory states also depend 
on internal parameters and on the number of links that the brain 
establishes  with the external world. These results agree with the 
physiological observations of the dynamic formation of neural circuitry 
which grows as brain develops and relates to external world.

\vspace{4.5cm}

*This paper is dedicated to Professor Karl H. Pribram in the occasion of 
his 80th birthday.

\newpage

\section{Introduction}

Since Lashely's experimental work in the forties it has been known that
many functional activities of the brain cannot be directly related to 
specific neural cells, rather they involve extended regions of the brain. 
In Lashely's words, as reported by Pribram\cite{P2}, "all behaviour seems 
to be determined by 
masses of excitation, by the form or relations or proportions of 
excitation within general fields of activity, without regard to particular 
nerve cells". Pribram's subsequent work, confirming and 
extending Lashely observations, brought him in the sixties to introduce 
concepts of Quantum Optics, such as holography, in brain 
modeling\cite{P2,PR}. The description of non-locality of brain functions, 
especially of memory storing and recalling, also was the main goal of the 
quantum brain model proposed in the 1967 by Ricciardi and 
Umezawa\cite{UR}. This model is based on the Quantum Field Theory (QFT) of 
many body systems and its main ingredient is the mechanism of spontaneous 
breakdown of symmetry. In recent years, the Ricciardi and Umezawa  quantum 
model, further developed by Stuart, Takahashi and Umezawa\cite{S1,S2} (see 
also \cite{CH}), has attracted much interest since it exhibits interesting 
features also related with the r\^ole of microtubules in the brain 
activity\cite{P2,PR,YA}. It has been shown that in the quantum brain model 
an essential r\^ole is played by the electrical dipole vibrational modes, 
from now on named dipole wave quanta (dwq), of the water molecules and of 
other biomolecules present in the brain structures\cite{YA}. Moreover, the 
extension of the model to dissipative dynamics has revealed to be crucial 
in order to allow a huge memory capacity\cite{VT}. The dissipative 
quantum model of brain has been recently investigated\cite{PV} also in 
relation with the possibility of modeling neural networks exhibiting 
collective dynamics and long range correlations among the net units. The 
study of quantum dynamical features for neural nets is of course of great
interest either in connection with computational neuroscience, either in
connection with quantum computational strategies based on quantum
evolution (quantum computation)\cite{QC}. 

In ref. \cite{VT} it has been considered the case  of
time independent frequencies  associated to the dwq. A more general
case is the one of time-dependent frequencies. The dwq may in fact undergo a number of fluctuating
interactions and then their characteristic frequency
may  accordingly change in time. The aim of this paper is to show that 
dissipativity and frequency 
time-dependence lead to the dynamical organization of the memories in 
space (i.e. to their localization in more or less diffused regions of the 
brain) and in time (i.e. to 
their longer or shorter life-time). 

The results we obtain appear to agree with physiological observations 
which show that many 
functions related to external world, such as movement, memory, vision, 
cannot be attributed to specific brain cells or structures, rather they 
involve different assembly of neurons in a way to "generate a connected 
whole"\cite{GRE}. The neural connectivity is observed to grow as the brain 
develops and relates to the external world. In previous studies of the 
quantum brain model the 
region involved in memory recording (and recalling) was extending to the 
full system. According to the results below presented, we have now 
"domains" of finite size involved in certain functions (essentially 
memory related functions). As observed in ref. \cite{VT}, the finiteness 
of the size of the domains implies an effective non-zero mass for the dwq 
(which, on the contrary, are 
massless in the case of the infinite volume system) and this in turn 
implies the appearance of related time-scales for the dwq propagation 
inside the domain. The arising of these time-scales seems to match 
physiological observation of time lapses observed in gradual recruitment 
of neurons in the establishment of brain functions\cite{GRE}. On the other 
hand, frequency time-dependence also introduces a fine structure in the 
decay behavior of memories, which adds up to the memory decay features 
implied by dissipativity (see ref. \cite{VT}).

The paper is organized as follows. In Sec. 2, for the reader convenience,
we present a summary of the main features of the dissipative quantum 
brain model.
In Sec. 3 we show that such a couple of equations 
may actually be derived from the spherical Bessel equation of given 
order, say 
$n$, when the frequency is (exponentially) time-dependent. We will see 
that the order $n$ provides a measure of the "openness" of the brain over 
the 
external world, i.e. of the number of links (couplings) that the brain  
can establish with the environment.
In Sec. 4 we consider the emergence of finite life-times for the memories 
and relate them to the sizes of the corresponding domains in 
the brain. Sec. 5 is devoted to further remarks and to conclusions. 
Finally, some details of the 
mathematical formalism are reported in the Appendix.

\section{The dissipative quantum brain model}

As it is well known, in quantum field 
theory (QFT) the spontaneous breakdown of symmetry occurs when the 
Lagrangian is invariant under certain 
group of continuous symmetry, say $G$, and the vacuum or ground state of 
the system 
is not invariant under $G$, but under one of its subgroups, say $G'$. In 
such a case, the ground state exhibits observable ordered patterns, 
specific of the breakdown of $G$ into $G'$. It can be shown \cite{IT,U} 
that these observable ordered  
patterns are generated by the coherent condensation in 
the ground state of massless quasiparticles called Nambu-Goldstone (NG) 
modes. These are thus the carriers of the 
ordering information in the ground state or vacuum. Their range of 
propagation covers the 
full system and therefore they manifest themselves as collective 
modes (we are assuming to be in the infinite volume limit as 
required by QFT).
The observable specifying the degree of ordering of the vacuum (called the
order parameter) acts as a macroscopic variable for the system and 
is specific of the kind of symmetry into play. Its 
value is related with the
density of condensed NG bosons in the vacuum.
Such a value may
thus be considered as the {\it code} specifying the vacuum of
the system (i.e. its physical phase) among many possible degenerate 
vacua. 

In the quantum model of brain\cite{UR}-\cite{YA} the information storage 
function, i.e. memory recording, is
represented by the "coding" of (i.e. the ordering induced in) the ground 
state by means of the
condensation of NG modes due to the breakdown of 
the rotational 
symmetry of the electrical dipoles of the water molecules. The 
corresponding NG mode are the vibrational dipole wave quanta (dwq) 
(in analogy to what happens in 
the QFT modeling of living matter\cite{DG}).
The trigger of the symmetry breakdown is the external 
informational input. It has been shown\cite{VT} that by taking into 
account the intrinsic dissipative character of the brain dynamics, namely 
that the brain is an {\it open system} 
continuously {\it linked} (coupled) with the environment, the memory 
capacity can be enormously enlarged, thus solving one of the main problems 
left unsolved in the original formulation of the quantum brain model.

The procedure of the canonical quantization of 
a dissipative system requires\cite{VT,QD} the
"doubling" of the degrees of freedom of the system.  Let
$A_{\kappa}$  and ${\tilde A}_{\kappa}$ denote the dwq  mode and its
"doubled mode", respectively. The suffix $\kappa$ denotes the momentum. 
In general there could be other suffices denoting other degrees of 
freedom of the modes $A$ and $\tilde A$. The  $\tilde A$ mode is the 
"time-reversed image", i.e. the "mirror in time image"  of the $A$ mode 
and it 
represents the environment mode.
Quantum field modes are associated with oscillator modes in the canonical 
formalism of QFT. In the present dissipative case the
$A_{\kappa}$  and ${\tilde A}_{\kappa}$ modes are associated 
with {\it damped} oscillator modes and their time-reversed image, 
respectively. 
The 
"couple" of the (classical) equations to be quantized is thus
\bea
\stackrel{..}u\,+\,L \stackrel{.}u\,+\, \omega^{2} u\,=
&\,0 ~,
\nonumber \\
\stackrel{..}v\,-\,L \stackrel{.}v\,+\, \omega^{2} v\,=
&\,0 ~,
\lab{pm3}\eea 
where the $u$ and $v$ variables 
are related to the $A_{\kappa}$  and 
${\tilde A}_{\kappa}$ (see refs. \cite{VT} and \cite{QD} for details).
Notice that we are here considering time-independent frequencies. 

Let
${\cal N}_{A_{\kappa}}$ and ${\cal N}_{{\tilde A}_{\kappa}}$  denote the
number of ${A_{\kappa}}$ and ${\tilde A}_{\kappa}$ modes,
respectively. 
Taking into account dissipativity requires that the memory state,
denoted by the vacuum ${|0>}_{\cal N}$ , is a condensate of {\it
equal number} of $A_{\kappa}$  and ${\tilde A}_{\kappa}$ modes, for any
$\kappa$ : such a requirement ensures in fact that the flow of the
energy exchanged between the system and the environment is
balanced. Thus, the difference between the number of tilde and non-tilde
modes must be zero:  ${\cal N}_{A_{\kappa}} - {\cal N}_{{\tilde
A}_{\kappa}} = 0$, for any $\kappa$.  The label ${\cal N}$ in
the vacuum symbol ${|0>}_{\cal N}$  specifies  the set of integers
$\{{\cal N}_{A_{\kappa}}, ~for~any~  \kappa \}$ which indeed defines the
"initial value" of the condensate,  namely the {\it code} number
associated to  the information recorded at time $t_{0} = 0$. 

Note now
that the requirement  ${\cal N}_{A_{\kappa}} - {\cal N}_{{\tilde
A}_{\kappa}} = 0, $ for any $ \kappa$,  does not uniquely fix the code 
${\cal N} \equiv \{{\cal N}_{A_{\kappa}}, ~for~any~  \kappa \}$. Also  
${|0>}_{\cal N'}$
with ${\cal N'} \equiv \{ {\cal N'}_{A_{\kappa}};  {\cal
N'}_{A_{\kappa}} - {\cal N'}_{{\tilde A}_{\kappa}} = 0,  ~for~any~
\kappa \}$ ensures the energy flow balance and therefore also
${|0>}_{\cal N'}$ is an available memory state: it will correspond
however to a different code number  $(i.e. {\cal N'})$  and therefore to
a different information than the one of code ${\cal N}$. Thus,
infinitely many memory (vacuum) states, each one  of them corresponding
to a different  code $\cal N$, may exist: A huge number of sequentially
recorded informations may  {\it coexist} without   destructive
interference since infinitely many vacua  ${|0>}_{\cal N}$, for all
$\cal N$,  are {\it independently} accessible   in the sequential
recording process.  The "brain  (ground) state" may be represented as
the collection (or the  superposition) of the full set of memory states
${|0>}_{\cal N}$, for all  $\cal N$.
 
In summary, one may think of the brain as a complex system with a huge
number of   macroscopic states (the memory states). The degeneracy
among the vacua    ${|0>}_{\cal N}$ plays a crucial r\^ole in solving
the problem of memory capacity.  The  dissipative dynamics introduces
$\cal N$-coded "replicas" of the system   and information printing can
be  performed in each replica without destructive interference with
previously recorded informations in other replicas. A huge memory 
capacity is thus achieved. 

According to the original  quantum brain model\cite{UR},  the recall
process is described as the excitation of  dwq modes  under an external
stimulus which is  "essentially a replication signal" of  the one
responsible for   memory printing. When dwq are excited   the brain
"consciously feels"  the presence of the   condensate pattern in the
corresponding coded vacuum.  The replication signal thus acts as a probe
by which the brain  "reads" the printed information. Since the
replication signal   is represented in terms of ${\tilde
A}$-modes\cite{VT} these modes act   in such a reading as the "address"
of the information to be recalled.

In the next section we will show how the couple of eqs. (\ref{pm3}) may be 
derived from the spherical Bessel equation of order $n$ in the case of 
(exponentially) 
time-dependent frequency. Such a result is by itself interesting  since 
it points to the possibility of using the mathematically powerful tools 
provided by the theory of special functions in modeling brain 
functioning.

\section{The Bessel equation and dissipation}

For sake of full generality one should also
consider time-dependent frequencies. Besides the rather formal motivation 
of generality, the physical motivation which suggests 
to consider time-dependent frequencies is that these are of course more 
appropriate to realistic
situations where the dwq may undergo a number of fluctuating
interactions with other quanta and then their characteristic frequency
may accordingly change in time. In order to study the case of 
time-dependent frequencies we consider the spherical Bessel equation of 
order $n$ whose solutions constitutes a complete 
set of (parametric) decaying functions\cite{Abram,Smirnov,Jackson}:
\be
z^2 \ {d^{2} \over {dz}^{2}}M_{n} + 2z \ {d \over {dz}}M_{n} + [z^2  - n(n+1)] 
\ 
M_{n} \ = \ 0 ~.
\lab{qm42}\ee
Here $n$ is integer or zero. In the present section, for notational 
simplicity we will 
omit the suffix $k$ in all the considered quantities. We will restore 
it in the next section. As it is well known, Eq. (\ref{qm42}) 
admits as particular solutions the first and second kind Bessel 
functions, or their linear combinations (the Hankel functions).

Note that both $M_{n}$ and $M_{-(n+1)}$ are solutions of the same eq. 
(\ref{qm42}). By using the substitutions: $w_{n,l}= M_{n}\ \cdot
(x_{n})^{-l}$, $z =\epsilon_{n} x_{n}$ and $x_{n} =  e^{-t
/\al_{n}}$, 
where $\epsilon_{n}, \al_n$ are arbitrary 
parameters, eq.(\ref{qm42}) goes into the 
following one:
\be
\stackrel{..}w_{n,l} \ - \ \frac{2l+1}{\al_{n}}\ \stackrel{.} w_{n,l} + 
\left[\frac{l(l+1)-n(n+1)}{\al^{2}_{n}} \ +\   ( 
\frac{\epsilon_{n}}{\al_{n}})^{2}\ e^{-2t/\al_{n}} \right] \ w_{n,l}\ 
= \ 0,
\lab{qm7}\ee
where $\stackrel{.} w$ denotes derivative of $w$ with respect to time. 
Making the choice $l(l+1)=n(n+1)$ the degeneracy between the solutions 
$M_{n}$ and $M_{-(n+1)}$ is removed and two different equations are 
obtained,
one for $l=n$ and the other one for $l=-(n+1)$: 
\bea
&\stackrel{..}{ w}_{n,-(n+1)} \ + \ \frac{2n+1}{\al_{n}}\ 
\stackrel{.}{ w}_{n,-(n+1)} + 
[( \frac{\epsilon_{n}}{\al_{n}})^{2}\ e^{-2t/\al_{n}} ] \ w_{n,-(n+1)}\ = 
\ 0,
\nonumber\\
&\stackrel{..}{ w}_{n,n} \ - \ \frac{2n+1}{\al_{n}} \ 
\stackrel{.}{w}_{n,n} + 
[( \frac{\epsilon_{n}}{\al_{n}})^{2}\ e^{-2t/\al_{n}} ] \ w_{n,n}\ = \ 0 
~.
\lab{qm6}\eea
By setting $u \equiv { w}_{n,-(n+1)},$ and $v \equiv { w}_{n,n}$, and 
choosing
the arbitrary parameters $\al_{n}$ and $\epsilon_{n}$ in such a way that 
$\frac{2n+1}{\al_{n}} \equiv L$ and 
$\frac{\epsilon_{n}}{\al_{n}} \equiv \om_0$ do not depend on $n$ 
(and on time), we see that 
eq. (\ref{qm6}) are nothing else than the couple of eqs. (\ref{pm3}) of 
the 
dissipative model {\em with time-dependent frequency} $\omega_{n}(t)$:
\bea
\stackrel{..}u\,+\,L \stackrel{.}u\,+\, {\omega_{n}}^{2}(t) u \,=
&\ 0,
\nonumber \\
\stackrel{..}v\,-\,L \stackrel{.}v\,+\, {\omega_{n}}^{2}(t) v \,=
&\ 0,
\lab{qm3}\eea 
\be 
  \omega_{n}(t)\ =\ {\omega}_{0} \ e^{-\frac{L 
t}{2n+1}}.
\lab{qm4} \ee
We see that 
$\omega_{n}(t)$ approaches to the time-independent 
value $\om_{0}$ for $n \rightarrow \infty$: the 
frequency time-dependence is thus "graded" by $n$. $L$ and $\om_{0}$ are 
characteristic 
parameters of the system.

We note that $\al_{-(n+1)}=-\al_{n}$ and that the transformation 
$n \rightarrow -(n+1)$ leads to solutions (corresponding to $M_{-(n+1)}$) 
which we will not consider since they have frequencies which are 
exponentially increasing in time (cf. eq. (\ref{qm4})). These solutions 
can be 
respectively obtained from the ones of eq. (\ref{qm3}) by time-reversal $t 
\rightarrow -t$.

We finally note that the functions $w_{n,n}$  and $w_{n,-
(n+1)}$ are "harmonically
conjugate" functions in the sense that they may be reconducted to 
the single parametric oscillator 
\be
\stackrel{..}{r}_{n} +\Omega_{n} ^2(t)r_{n}=  0 ~,
\lab{qmm5}
\ee 
where $w_{n,-(n+1)}(t) = 
{1\over {\sqrt 2}} r_{n}(t)e^{{-L t\over 2}}$, \quad
 $w_{n,n}(t) = {1\over {\sqrt 2}} r_{n}(t)e^{{L t\over2}}$. $\Omega_n$
is the common frequency:
\be 
  \Omega_{n}(t) = \left [ \left ({\omega}_{n}^{2}(t)  - {{L  
  ^{2}}\over{4}} \right ) \right ]^{1\over{2}}.
  \lab{qm5} \ee

The quantization of the system (\ref{qm3}) can be now performed along the 
same 
line presented in refs. (\cite{QD,PLA,double}) and we summarize the main 
steps of the procedure in the Appendix.
We remark that the main features of the dissipative quantum model with 
constant frequency also hold in the present case of time-dependent 
frequency. We analyze in 
the next section the consequences on the memory recording 
process of the time-dependence of $\Omega_n$. As we will see, we will get 
some information on the size and on the life-time
of the memory domains.

\section{Life-time and localizability of memory states}

The meaning of the time-dependence of the frequency $\Omega_n$ of the 
coupled systems $A$ and $\tilde A$ represented by eq. (\ref{qmm5}) 
is that 
energy is not conserved in time and thus that the $A- {\tilde A}$ system 
does not really constitute a "closed" system, it does not really represent 
the "whole". On the other hand, we see from eq.(\ref{qm4}) that as $n 
\rightarrow \infty$, $\Omega_n$ approaches to a time independent quantity, 
which means that energy is conserved in such a limit, i.e. the $A- {\tilde 
A}$ system gets "closed" as $n \rightarrow \infty$. In other words, in the 
limit $n \rightarrow \infty$ 
the possibilities of the system $A$ to couple to ${\tilde A}$ (the 
environment) are "saturated": the system $A$ gets {\it fully} coupled to 
$\tilde A$. This suggests that $n$ represents the 
number of {\it links} between $A$ and ${\tilde A}$. When $n$ is not very 
large (infinity), the system $A$ (the brain) has not fulfilled its 
capability 
to establish links with the external world.

We observe now that in order the memory recording may occur the frequency
(\ref{qm5}) has to be real. Such a reality 
condition is found to be satisfied only in a definite span of
time, i.e., upon restoring the suffix $k$, for times $t$ such that 
$0 \le  t \le T_{k,n}$, with $ T_{k,n}$ 
given by
\be
T_{k,n}\,=\,\frac{2n+1}{L}\ \ln\left(\frac{2\om_{0,k}}{L}\right). 
\lab{qm8}\ee
Thus, the  memory recording processes can occur in limited time intervals 
which have  
$T_{k,n}$ as the upper bound, for each $k$. For times greater than 
$T_{k,n}$ memory recording cannot occur. Note that, for fixed $k$, 
$T_{k,n}$ grows linearly in $n$, which means that the time span useful for 
memory recording (the ability of memory 
storing) grows as $n$ grows, i.e. it grows as the number of links which 
the 
system is able to lace with the external world grows: more the system is 
"open" to the external world (more are the links), better it can memorize 
(high ability of learning). Also notice that the quantities $\om_{0}$ and 
$L$ are intrinsic 
to the system, they are {\it internal} parameters and may represent 
(parameterize) subjective attitudes.
 
We observe that our model is not able to provide a dynamics for the 
variations of $n$, thus we cannot say if and how and under which specific 
boundary conditions $n$ increases or 
decreases in time. However, the 
processes of learning are in our model each other independent, so it is 
possible, for example, that the ability in information recording may be 
different under different circumstances, at different ages, and so on. In 
any case, a higher or lower {\it 
degree of openness} (measured by $n$) to the external world may produce a 
better or worse ability in learning, respectively 
(e.g. during the childhood or the older ages, respectively). 

Coming back to $T_{k,n}$, we can now fix $n$ and analyze its behavior in 
$k$. We remark that  
${\Omega_{k,n}}^{2}(t) \ge 0$ at any given $t$ (the reality condition) 
implies   $k \ge {\tilde k}(n,t) \equiv k_{0} e^{{{L \over {2n+1}}t}}$, with $ k_{0} 
\equiv {L  
  \over  {2c}}$ at any given $t$ (note that $\omega_{0}= kc$). Thus we see that a threshold exists for 
the
$k$ modes of the memory process. We also note that such a kind of 
"sensibility" to external stimuli only depends on the internal parameters. 
On the other hand, the previous condition
on $k$ becomes less strict as $n$ grows (the threshold is lowered and a 
richer spectrum of 
$k$ modes is involved in the memory process). 

Another way of reading the above condition on $k$ is that it excludes 
modes
of wave-length greater than ${\tilde \lambda} \propto 
{1\over {{\tilde k}(n,t)}}$ for any given $n$ and $t$. The 
reality condition thus acts as an 
intrinsic infrared cut-off, which means that infinitely long 
wave-lengths (infinite volume limit) are 
actually precluded, and thus transitions through 
different vacuum states (which would be unitarily inequivalent vacua in 
the 
infinite volume limit) at given $t$'s 
are possible. This opens the way to both the possibilities, of 
"association of memories" and of "confusion" of memories (see also ref. 
\cite{VT}). We also remark that the existence of 
the ${\tilde \lambda}(n,t)$ cut-off means that (coherent) domains of sizes 
less 
or equal to ${\tilde \lambda}$ are involved in the memory 
recording, and that such a cut-off shrinks in time for 
a given $n$. On the other hand, a grow of $n$ opposes to such a shrinking. 
These cut-off changes correspondingly reflect on the memory domain sizes.

  \begin{figure}[t]
  \caption{``Lives'' of $k$ modes, for growing $k$}
\epsfig{file=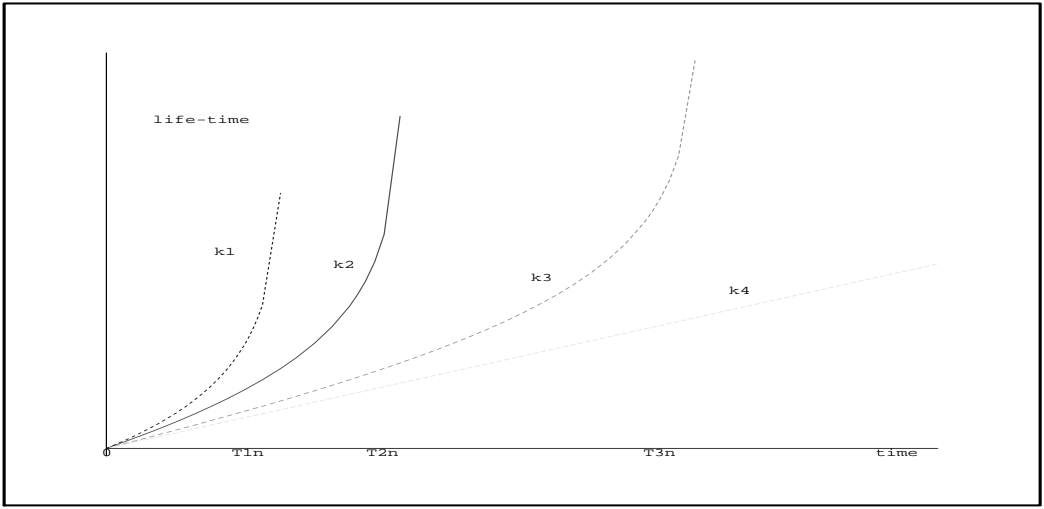,width=0.5\linewidth,height=1.0\linewidth,angle=-90}  
  %\caption{}
  \end{figure}
  %%%%%

We can estimate the domain evolution by introducing the quantity 
$\Lambda_{k,n}(t)$\cite{PLA, double}: 
  \be 
   e^{-2\Lambda_{k,n}(t)} \ =\  
  \frac{e^{-t \frac{L}{2n+1}}\,
{\rm sinh}\frac{L}{2n+1}({\rm T}_{k,n}-t)}{{\rm sinh} 
  \frac{L}{2n+1}{\rm T}_{k,n}}, \qquad \, \Lambda_{k,n}(t) \ 
\ge 0, \,{\rm   for \; any \;~ t} ~,
  \lab{c6}\ee 
i.e. $\Lambda_{k,n}(0)=0$ for any $k$, and 
$\Lambda_{k,n}(t)\rightarrow\infty$ for
$t \rightarrow\ T_{k,n}$ for any given $n$.
Then $\Omega_{k,n}$ may 
be expressed in the form: 
\be {\Omega_{k,n}(\Lambda_{k,n}(t))}\,=\,  \Omega_{k,n}(0) 
e^{-\Lambda_{k,n}(t)}~. 
  \lab{c5}\ee 
\begin{figure}[t]
  \caption{``Lives'' of $k$ modes, for growing $n$ and fixed $k$}
  \epsfig{file=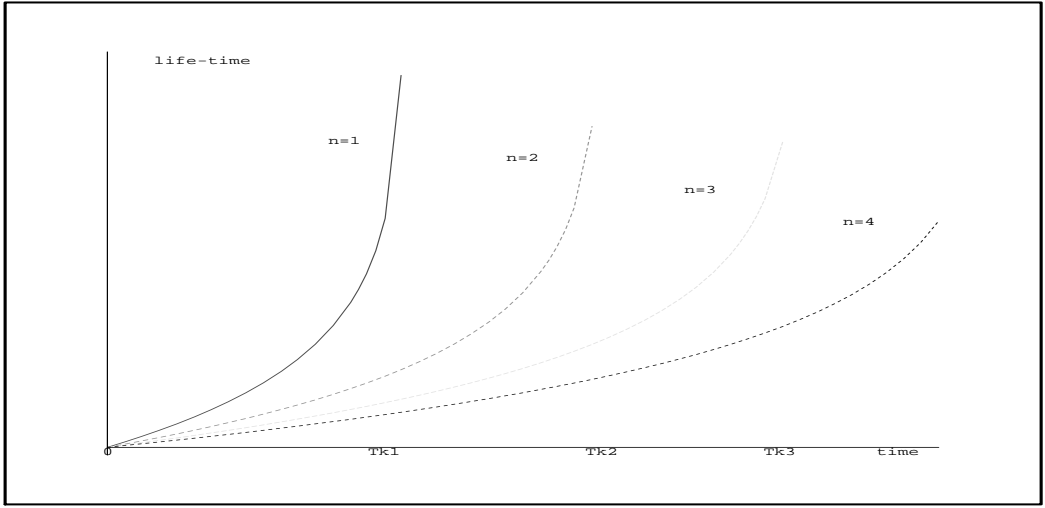,width=0.5\linewidth,height=1.0\linewidth,angle=-90}
 %\caption{}
 \end{figure}
\begin{figure}[h]
 \caption{``Lives'' of $k$ modes, for growing $n$ and `small` $k$  }
  \epsfig{file=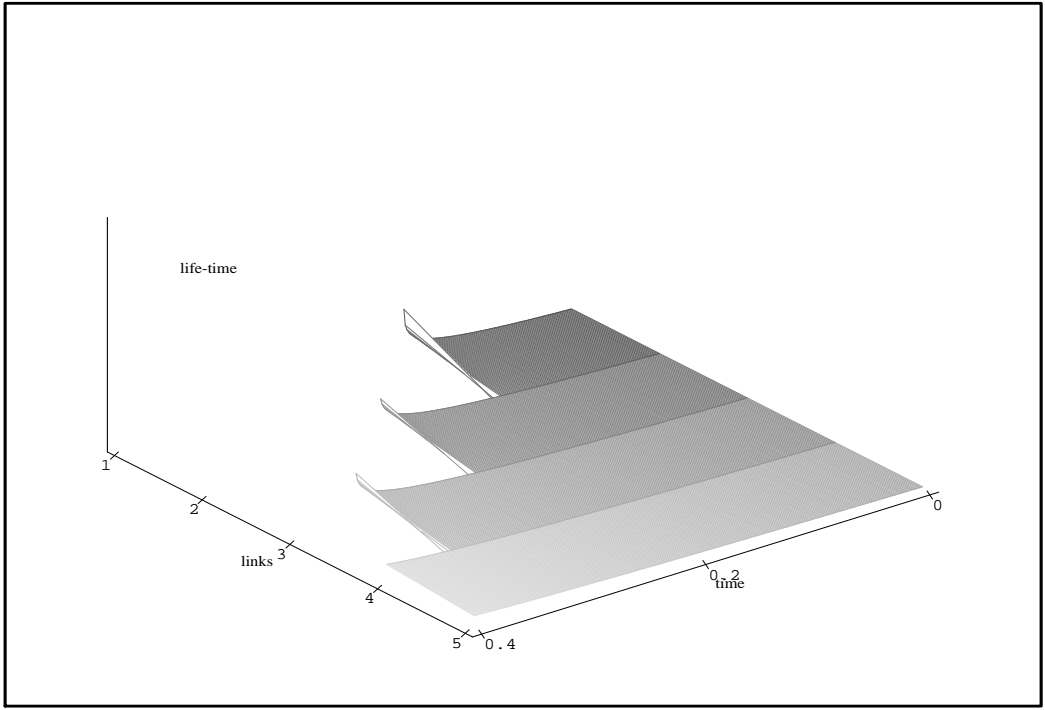,width=0.7\linewidth,height=1.0
\linewidth,angle=-90}
%%\caption{}
\end{figure}
\begin{figure}[h]
 \caption{``Lives'' of $k$ modes, for growing $n$ and 
`big` $k$  }
  \epsfig{file=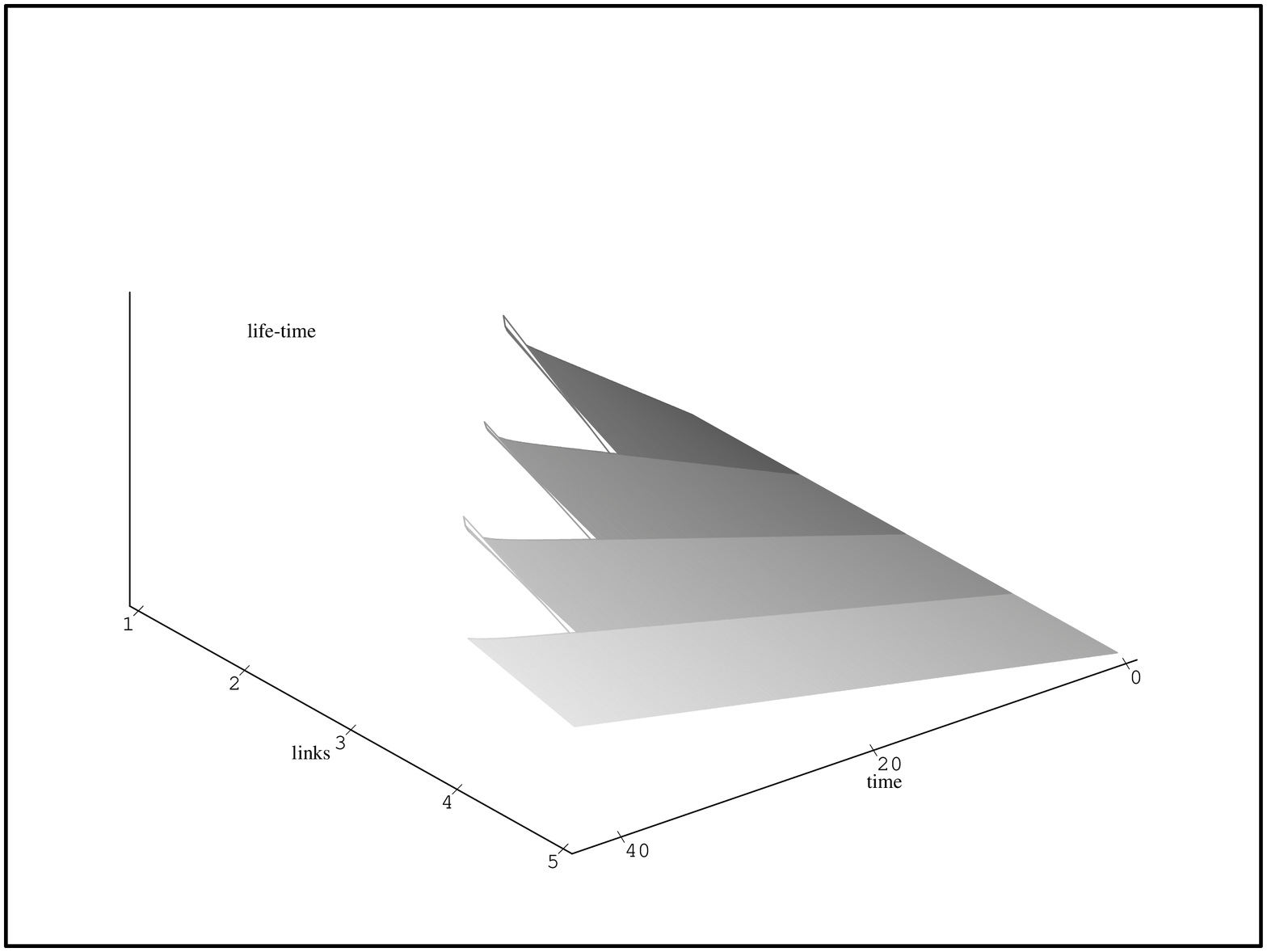,width=0.7\linewidth,height=1.0\linewidth,angle=-90}
%  %\caption{}
\end{figure}
Since ${\Omega_{k,n}(\Lambda_{k}({\rm T}_{k,n}))}\,=\,0$, we 
see that $\Lambda_{k,n}(t)$ acts as a life-time, say $\tau_{k,n}$, with 
$\Lambda_{k,n}(t) \propto \tau_{k,n}$,
for the mode $k$.
Modes with larger $k$ have a "longer" life with reference 
to time $t$. 
In other words, each $k$ mode "lives" with a proper time $\tau_{k,n}$, so 
that the mode is born when $\tau_{k,n}$ is zero and it dies for 
$\tau_{k,n}$ 
$\rightarrow$ $\infty$. 

The "lives" of the $k$ modes, for growing $k$, are sketched in the figures 
1-4. In Fig. 1 the lives are drawn for growing $k$ and fixed $n$ ($n = 
1$). Fig. 2 shows the "lives" of a mode of given $k$ for different values 
of $n$. The $\Lambda_{k,n}$s are drawn versus $t$, 
reaching the blowing up values in correspondence of the abscissa points 
$T_{k,n}$. In Fig. 1 the values of $k$ have been chosen so that $k_{1}$ is 
one tenth of $k_{3}$ and $k_{2}$ is one tenth of $k_{4}$.
Only the modes satisfying the reality condition are present at certain 
time 
$t$, being the other ones decayed.
This introduces from one side an hierarchical organization of memories 
depending on their life-time. Memories with a specific spectrum of $k$ 
mode components 
may coexist, some of them "dying" before, some other ones persisting 
longer in dependence on the number in the spectrum of the smaller or 
larger $k$ components, 
respectively (see Fig. 2). On the other hand, as observed above, since 
smaller or larger $k$ modes 
correspond to larger or smaller waves lengths, respectively, we also see 
that the (coherent) memory domain sizes correspondingly
are larger or smaller. 

In conclusion, more persistent memories 
(with a spectrum more populated by the higher $k$ components) are also 
more "localized" than shorter term 
memories (with a spectrum more populated by the smaller $k$ components), 
which instead extend over larger domains. We have thus reached a "graded" 
non-locality for memories depending on $n$ and on the spectrum of their 
$k$ modes components, which is also related to their life-time or 
persistence. 

Figures 3 and 4 describe the cooperative r\^ole of $n$ and 
$k$ in supporting a long-lived memory: Fig. 3 shows the behavior of 
small-$k$ (with respect to $k_{0}$)modes for growing $n$ and 
Fig. 4 shows the behavior of large-$k$ (with respect to $k_{0}$) modes for growing $n$. It is evident that the time of memory 
persistence (expressed in adimensional units) increases of several orders 
of magnitude as the value of $k$ grows. 

Since $k$ denotes the momentum of the dwq $A$ and of the $\tilde A$ 
excitations, it is expected that, for given $n$, "more impressive" is the 
external stimulus (i.e. stronger is the coupling with the external world) 
greater is the number of high $k$ momentum excitations produced in the 
brain and more "focused" is the "locus" of the memory.

\section{Final remarks and conclusions}
 
In the quantum model of brain memory recording is obtained by coherent 
condensation of the dipole wave quanta in the system ground state or 
vacuum. In the non-dissipative case the memory states are thus stable 
states (infinitely long-lived states): there is no possibility to forget. 
On the contrary, in the dissipative  case the memory states have finite 
(although long) life-times\cite{VT}. Above we have seen that a rich 
structure in the life-time behavior of single modes of momentum $k$ is 
implied by the frequency time-dependence of the dwq. Such a fine structure 
determines to some extend the memory storing ability and it appears to 
depend also on the brain internal parameters. There is therefore a 
specificity of the system which enters in determining its functional 
features, as of course it is natural to be. Moreover, this same fine 
structure 
also depends on the {\it degree of openness} of the brain, i.e. on the 
number of {\it links} which it is able to establish with the external 
world. On the other hand, the number of such links also  
turns out to be relevant to the memory recording ability. 

Although at this stage our model does not give us quantitative 
predictions, nevertheless the qualitative behaviors and results above 
presented appear to fit well with the physiological observations\cite{GRE} 
of the formation of 
connections among neurons as a consequence of the establishment of the 
links between the brain and the external world. More the brain relates to 
external environment, more neuronal connections will form. 
Connections appear 
thus more important in the brain functional development than the single 
neuron activity. Here we are referring to 
functional or effective connectivity, as opposed to the structural or 
anatomical one\cite{GRE}. The last one can be described as 
quasi-stationary. The 
former one is highly dynamic with modulation time-scales 
in the range of hundreds of milliseconds\cite{GRE}. Once these functional 
connections are formed, they are not necessarily fixed. On the contrary, 
they may quickly change in a short time and new configurations of 
connections may be formed extending over a domain including a larger or a 
smaller number of neurons. Such a picture finds a possible description in our 
model, where the coherent domain formation, size and life-time depend on 
the number of links that the brain sets with its environment and  
on internal parameters.
As shown elsewhere (see ref. 
\cite{VT} and \cite{DG}) the em field propagates in ordered domains 
in a self-focusing fashion, thus producing an highly dynamic net of 
filaments and tubules which may model the highly dynamic neuronal 
connection assembly and disassembly. 
The prerequisite for the connection formation is the above described
dynamic generation of ordered domains of dwq.

There is a further remarkable aspect in the occurrence of finite size 
domains. 
The finiteness of the domain size spoils the unitary inequivalence among 
the vacua of the domain. Then, in 
the case of open systems,  transitions among
(would be) unitary inequivalent vacua may  occur (phase transitions) for
large  but  finite  volume,  due  to  coupling  with  external
environment.It has been shown that small perturbation may drive the system 
through its vacua\cite{CGV}. In this way occasional (random) 
perturbations may play an important r\^ole in the complex behavior of the 
brain activity.
However, transitions among different vacua may be not a
completely negative feature of the model: once transitions are allowed, 
the possibility of associations of memories
("following a path of memories") becomes possible. Of course, these
"transitions" should only be allowed up to a certain degree in order to
avoid memory "confusion" and difficulties in the process of storing
"distinct" memories. The opposite situation of strict inequivalence 
among different vacua (in the case of very large or infinite size domain)
would correspond to the absence of any "transition" among memory states 
and thus to "fixation" or "trapping" in some specific memory state.

In connection with the recall mechanism, we note (see ref. \cite{VT}) 
that the dwq  acquires an effective non-zero mass due to the domain finite 
size. Such an effective mass will then acts as a
threshold in the excitation energy of dwq so that, in order to trigger
the recall process  
an energy supply equal or greater than such a
threshold is required. When the energy supply is lower than the required
threshold a "difficulty in recalling" may be experienced.  At the same
time, however, the threshold may positively act as a "protection"
against unwanted perturbations (including thermalization) and
cooperate to the stability of the memory  state.  In the case of zero
threshold (infinite size domain) any replication signal could excite  the 
recalling  and the brain would fall in a state of "continuous flow of   
memories".

We further observe that the differences in the life-time of the $k$ 
components may produce the corruption of the spectral structure of the 
memory information (of the memory code) with consequent more or less 
severe memory "deformations".
On the other hand, due to the dissipation, at some time $t = \tau$,   
conveniently larger than the memory life-time,
the memory state   $|0>_{\cal N}$ is reduced to the "empty" vacuum
$|0>_{0}$ where  ${\cal N}_{\kappa} = 0$ for all $\kappa$:  the
information has been {\it forgotten}.  At the time $t = \tau$  the state
$|0>_{0}$ is available for recording a new information. In order to not 
completely forget  certain information,  one needs  to "restore" the 
${\cal N}$ code, namely to "refresh" the memory by {\it brushing up} the 
subject (external stimuli maintained memory). 

Finally, we note that one is actually obliged to consider the dissipative, 
irreversible time-evolution: in fact {\it after} information has been 
recorded, the brain state is completely
determined and the brain cannot be brought to the state 
in which it was {\it before} the information printing occurred.  Thus,
the same fact of getting information introduces {\it the arrow of
time} into brain dynamics. In other words, 
it introduces a partition in the time evolution, 
namely the {\it distinction} between the past and the future, a
distinction which did not exist {\it before} the information recording.
It can be shown that dissipation and the frequency 
time-dependence imply 
that the evolution of the memory state is controlled
by the entropy variations\cite{VT}: this feature reflects indeed the
irreversibility of time evolution (breakdown of time-reversal symmetry). 
The stationary condition for the free energy functional leads then to 
recognize the memory state  $|0(t)>_{\cal N}$ to be a finite temperature
state\cite{U}, which opens the way to the possibility of thermodynamic 
considerations in the brain activity.
In this connection we observe that the ``psychological arrow of time'' which 
emerges in the brain dynamics turns 
out point in the same direction of the ``thermodynamical arrow of time'', which points in the increasing entropy 
direction. It is interesting to note that both these arrows, the 
psychological one and the thermodynamical one, also point in the same 
direction of the "cosmological arrow of time" defined by the expanding 
Universe direction\cite{double, Hawking}. On the subject of the 
concordance of the 
three different arrows of time there is an interesting debate still going 
on(see e.g. \cite{Hawking}). It is remarkable that the dissipative 
quantum model of brain let us reach a conclusion on the psychological 
arrow of time which we commonly experience.

This work has been partially supported by INFM and by MURST.

\section{Appendix}

We briefly summarize the main steps in the quantization procedure of the 
eqs. (\ref{pm3}). For more details see refs. \cite{QD,PLA,double} where, 
although in a different context, a complete treatment is given. In the 
following we will omit the suffices for simplicity, unless they are needed 
in order to avoid misunderstanding.
It turns out to be convenient to introduce 
the canonical transformations
\be u(t)\ =\  \frac{U(t) + V(t)}{\sqrt 2}, \qquad
v(t)\ = \ \frac{U(t) - V(t)}{\sqrt 2}.
 \lab{p35}\ee

The Hamiltonian for our coupled oscillator equations is: 
  \be {\cal H}   = {1 \over 2} {p_U}^2 + {1 \over 2}{\Omega}^2(t)U^2 
  -{1 \over 2} {p_V}^2  
    - {1 \over 2}{\Omega}^2(t)V^2   - 
  {\Gamma} ({p_U} V + {p_V} U).\lab{p42}\ee 
  with $\Gamma \equiv {L \over 2}$. We assume 
  $\Omega(t)$ to  
  be real (for any $k$ and any $t$) 
  in order to avoid 
  over-damped regime. As we have discussed in the text, 
  this condition acts as  
  a cut-off on $k$. 
  The conjugate momenta are 
  \be 
p_U =  \dot U + {L \over 2} V ~~,\qquad 
p_V = - \dot V -{L \over 2} U ~~.
\lab{p39}\ee 

 We introduce the annihilation  operators 
  \be 
  A  = \frac{1}{\sqrt{2}} \left(\frac{p_U}{\sqrt{\hbar\om_{0}}}-iU 
  \sqrt{\frac{\om_{0}}{\hbar}} 
  \right), \qquad {\tilde A} 
   = \frac{1}{\sqrt{2}} \left(\frac{p_V}{\sqrt{\hbar\om_{0}}}- 
   iV\sqrt{\frac{\om_{0}}{\hbar}} 
  \right),  
  \lab{p44}\ee 
  and the corresponding creation operators with usual commutation 
relations. Here $\om_{0}$ is an arbitrary constant frequency. Then it can 
be shown that the vacuum state $| 0>$ is unstable:
  \be 
  <0(t) | 0> \, \propto \exp{\left ( - t  \Gamma \right )} 
  \rightarrow 0 \; ~for~large~t , 
  \lab{p712}\ee  
  i.e. time evolution brings "out" of the initial-time Hilbert  
  space for large $t$. This compels us to work in the framework of QFT 
(not just of Quantum Mechanics!)\cite{PLA}. In order to set up  
  the formalism in QFT we have to consider the infinite volume limit;  
  however, as customary, we will work at finite volume and at the end of  
  the computations we take the limit $V \rightarrow \infty$. The QFT  
  Hamiltonian  is 
  \be 
  {\cal H}  = {\cal H}_{0} + {\cal H}_{I_1} + {\cal H}_{I_2} 
  \lab{ph1}\ee 
  \be 
  {\cal H}_{0} =  \sum_{ k} \half \hbar \Om_{0,{ k}}(t) 
  (A^{\dagger}_{ k} A_{ k} - {\tilde A}^{\dagger}_{ k} {\tilde A}_{ k} ) 
~, 
  \lab{ph2}\ee 
  \be 
  {\cal  H}_{I_1} = - \sum _{ k}{1\over 4}\hbar \Om_{1,{ k}}(t) 
  \left[\left(A_{ k}^{2} 
  + {A^{\dagger 2}_{ k}} \right) - \left( {{\tilde A}_{ k}}^{2} + 
{{\tilde A}^{\dagger 2}_{ k}} \right)\right] ~, 
  \lab{ph3}\ee 
  \be 
  {\cal H}_{I_2} = i\sum _{ k} {\Gamma}_{ k} 
  {\hbar} \left(A^{\dagger}_{ k} {\tilde A}^{\dagger}_{ k} -A_{ k}{\tilde 
A}_{ k}  
  \right) ~, 
  \lab{ph4}\ee 
  with  $\Omega_{0,1,k}= \om_{0}  
  \left(\frac{{{\Om}_k}^{2}(t)}{\om^{2}_{0}}\pm 1\right) ~~$ and 
  \be 
  [A_{ k},A^{\dagger}_{{ k}'}]  =  {\delta}_{{ k},{ k}'} =  
   [{\tilde A}_{ k},{\tilde A}^{\dagger}_{{ k}'}],\qquad    
  {[}A_{ k},{\tilde A}_{{ k}'}{]}  =  0\,=\,  
   [A^{\dagger}_{ k},{\tilde A}^{\dagger}_{{ k}'}] . 
  \lab{p51}\ee 
  We remark that 
  \be 
  [{\cal H}_{0}, {\cal H}_{I_2}] = 0 = [{\cal H}_{I_1} , {\cal H}_{I_2}]~, 
  \lab{p62}\ee 
  which guarantees that the minus sign appearing in  
  ${\cal H}_{0}$ is not 
  harmful, i.e., once one starts with a positive definite Hamiltonian it  
  remains lower bounded under time evolution.  
      
  By using the transformation ${\cal H} 
  \rightarrow {{\cal H}^{\prime}} \equiv S(\theta)^{-1} {\cal H}
   S(\theta)$ with $ S(\theta) \equiv \prod_{ k} 
  e^{-i\theta_{k}(t)K_{2,k}}$ and 

  \be 
  \tanh {\theta_{k}(t)} =  
  -{\Om_{1,k} (t) \over \Om_{0,k}  
  (t)} 
  \lab{c0}\ee 
  at any $t$ for any given $k$ we can  
   "rotate away" ${\cal H}_{I_1}$: 
  \be 
  {{\cal H}^{\prime}} \equiv S(\theta)^{-1} {\cal H} S(\theta)
   = {\cal H}^{\prime}_{0} + {\cal H}_{I_{2}}~~. 
  \lab{p74}\ee 
  
  Note that
  \be 
  {{\cal H}^{\prime}}_{0} = \sum_{ k}\hbar \Om_{ k}(t) 
  (A_{ k}^{\dagger} A_{ k} - {\tilde A}_{ k}^{\dagger} {\tilde A}_{ k} 
)~~,
  \lab{ph5}\ee 
  and $ [{{\cal H}^{\prime}}_{0} , {\cal H}_{I_{2}} ] = 0 $.
   When the initial state, say at arbitrary  
  initial time $t_{0}$, ($t_{0} =0$, ${\theta}_{k} ({0})\equiv 
{\theta}_{k}$  
  for sake of simplicity)  
  is the {\it vacuum} $|0>$ for ${{\cal H}^{\prime}}_{0}$, with 
  $A_{k} |0> = 0 =  
  {\tilde A}_{k}|0>$, 
  the state 
  $
  |0(\theta)> = S(\theta) |0> ~ 
  $  is the zero energy eigenstate 
  (the {\it vacuum}) of ${\cal H}_{0} + {\cal H}_{I_1}$ at $t_{0}$: 
  \be 
  ({\cal H}_{0} + {\cal H}_{I_1})|_{t_{0}}|0(\theta )> = S(\theta)
  {\cal H}^{\prime}_{0}|0> = 0. ~~  
  \lab{p77}\ee 
   
  We observe that the 
  operators $A_{k}$ and ${\tilde A}_{k}$ transform under 
  $\exp{\left ( -i{\theta}_{k} K_{2,k} \right )}$  as 
  \be 
  A_{k}  \mapsto A_{k} (\theta) =  {\it e}^{ 
  -i{\theta}_{k} K_{2,k} }A_{k}  {\it e}^{i{\theta}_{k} K_{2,k}} = 
  A_{k}  \cosh{(\half{\theta}_{k} )} + {A_{k}} ^{\dagger} \sinh{( 
  \half{\theta}_{k} )}~, 
  \lab{p7261}\ee
  \be
  {\tilde A}_{k}  \mapsto {\tilde A}_{k} (\theta)  =  {\it e}^{ 
  -i{\theta}_{k} K_{2,k} }{\tilde A}_{k}  {\it e}^{i{\theta}_{k} K_{2,k} } 
= 
   {\tilde A}_{k}  \cosh{(\half{\theta}_{k} )} + {{\tilde A}_{k}} 
^{\dagger} \sinh{( 
  \half{\theta}_{k} )}~. 
  \lab{p7262}\ee 
  One has  
  $ 
  A_{k} (\theta) |0(\theta)> = 0 = {\tilde A}_{k} (\theta) |0(\theta)> $~.
 
 The  
  $t$-evolution of the vacuum $|0(\theta)>$ is obtained as (at
  finite volume $V$):
  \be 
  |0(\theta,t)> = \prod_{ k} {1\over{\cosh{(\Gamma_{ k} t)}}} \exp{ 
  \left ( \tanh {(\Gamma_{ k} t)} J_{{ k}, +}(\theta) \right )}  
  |0(\theta)> \quad,  \lab{pq79}\ee 
  with $J_{k,+}(\theta) = A^{\dagger}_{k}(\theta)  {\tilde 
A}^{\dagger}_{k}(\theta)$. 
  We have  $A_{ k}(\theta,t) |0(\theta,t)> = 
  0 = {\tilde A}_{ k}(\theta,t) 
  |0(\theta,t)>$ with 
  \be
  A_{k}(\theta)\mapsto A_{k}(\theta,t)  = {\it e}^{- i {t\over{\hbar}} 
{\cal  
  H}_{I_2}} A_{k}(\theta) {\it e}^{i {t\over{\hbar}} {\cal H}_{I_2}}  =   
  A_{k}(\theta) \cosh{({\Gamma}_{k} t)} - {\tilde A}_{k}(\theta)^{\dagger} 
\sinh{( 
  {\Gamma}_{k} t)} \quad , 
  \lab{795}\ee 
  \be 
  {\tilde A}_{k}(\theta) \mapsto {\tilde A}_{k}(\theta,t)  = {\it e}^{- i 
{t\over{\hbar}} {\cal  
  H}_{I_2}} {\tilde A}_{k}(\theta) {\it e}^{i {t\over{\hbar}} {\cal 
H}_{I_2}}  =  -  
  A_{k}(\theta)^{\dagger} \sinh{({\Gamma}_{k} t)} + {\tilde A}_{ 
k}(\theta) \cosh{( 
  {\Gamma}_{k} t)} \quad. \lab{p796}\ee 
  Notice that these are the time-dependent, canonical Bogolubov 
  transformations.
  
  The state $|0({\theta},t)>$ is a normalized state,~ $ 
  <0(\theta,t) | 0(\theta,t)> = 1 \quad \forall t$, ~and  is  a  
  $su(1,1)$  generalized coherent state. 
 It can be shown that in the 
  infinite volume limit 
  \be 
  <0(\theta,t) | 0(\theta)> {\mapbelow{V \rightarrow \infty}} 0 \quad 
\forall  
  \, t \quad , 
  \lab{p713}\ee 
  \be 
  <0(\theta,t) | 0(\theta',t') > {\mapbelow{V \rightarrow \infty}} 0 \quad  
  with ~~ \theta' \equiv \theta (t_{0}'),~~\forall \, t\, , t'\, , t_{0}'  
  \quad , \quad t \neq t'~~ .  \lab{p714}\ee

  Eqs. (\ref{p713}) and (\ref{p714}) show that in the infinite volume
  limit the vacua at $t$ and at $t'$, for any $t$ and $t'$, are
  orthogonal states and thus the corresponding Hilbert spaces are 
unitarily inequivalent spaces. 

  The number of modes of type $A_{k}(\theta)$ in the state  
  $|0(\theta,t)>$ is given, at each instant $t$ by 
  \be 
  {N}_{A_{k}}(t) \equiv < 0(\theta,t) | A_{k}^{\dagger} 
  (\theta)  
  A_{ k}(\theta) | 0(\theta,t) > = 
   \sinh^{2}\bigl( \Gamma_{ k} t \bigr) \quad , 
  \lab{p274}\ee 
  and similarly for the modes of type ${\tilde A}_{k}(\theta)$.  
   
  We also observe that the
  number $\left( N_{A_{k}} - N_{{\tilde A}_{k}}  
  \right)\,$ is a constant of motion for any $ k$ and any $\theta$. 
  Moreover, one can show that 
  the creation of a mode $A_{k}(\theta)$ is equivalent to the 
destruction of a mode ${\tilde A}_{k}(\theta)$ and vice-versa.  
This means  
  that the ${\tilde A}_{k}(\theta)$ modes can be interpreted  as the {\sl  
  holes} for the modes $A_{k}(\theta)$: the ${\tilde A}$ system can be  
  considered as the sink where the energy dissipated  
  by the $A$ system flows. 

Finally thermal properties of the vacuum $| 0(\theta,t) >$ can be 
analyzed\cite{QD, PLA, double} 
and the $\tilde A$ modes appear to represent the thermal bath (the 
environment) modes.

\end{document}